\documentclass{aa}  

\usepackage{graphicx}
\usepackage{txfonts}
\AddToHook{begindocument/before}{\usepackage{hyperref}}
\begin{document}


%
%

   \title{Constraining the inner boundaries of COCONUT through plasma $\beta$ and Alfvén speed}


   \author{M. Brchnelova
          \inst{1}
          \and
          B. Gudiksen \inst{2, 3} \and
          M. Carlsson \inst{2, 3} \and
          A. Lani \inst{1} \and 
          S. Poedts \inst{1,4}
          }

   \institute{Centre for Mathematical Plasma Astrophysics, KU Leuven,
              Celestijnenlaan 200B, 3001, Leuven\\
              \email{michaela.brchnelova@kuleuven.be}
        \and  Rosseland Centre for Solar Physics, University of Oslo, PO Box 1029 Blindern, 0315 Oslo, Norway
        \and Institute of Theoretical Astrophysics, University of Oslo, PO Box 1029 Blindern, 0315 Oslo, Norway
        \and Institute of Physics, University of Mariia Curie-Sk{\l}odowska, ul.\ Radziszewskiego 10, 20-031 Lublin, Poland 
             }

   \date{Received ? ??, ????; accepted ? ??, ????}

 
  \abstract
   {Space weather modelling has been gaining importance due to our increasing dependency on technology sensitive to space weather effects, such as satellite services, air traffic and power grids. Improving the reliability, accuracy and numerical performance of space weather modelling tools, including global coronal models, is essential to develop timely and accurate forecasts and to help partly mitigate the space weather threat. Global corona models, however, require accurate boundary conditions, for the formulations of which we have very limited observational data. Unsuitable boundary condition prescriptions may lead to {inconsistent} features in the solution flow field and spoil the code's accuracy and performance. 
   }
   {In this paper, we develop an adjustment to the inner boundary condition of the COolfluid COrona uNstrUcTured (COCONUT) global corona model to better capture the dynamics over and around the regions of stronger magnetic fields by constraining the plasma $\beta$ and the Alfvén speed. 
   }
   {Using data from solar observations and solar atmospheric modelling codes such as Bifrost, we find that the baseline homogeneous boundary condition formulations for pressure and density do not capture the plasma conditions physically accurately. We develop a method to adjust these prescribed pressure and density values by placing constraints on the plasma $\beta$ and the Alfvén speed that act as proxies.
   }
   {We demonstrate that we can remove {inexplicable} fast streams from the solution by constraining the maximum Alfvén speed and the minimum plasma $\beta$ on the boundary surface. We also show that the magnetic topology is not significantly affected by this treatment otherwise. 
   }
   {The presented technique shows the potential to ease the modelling of solar maxima, especially removing {inexplicable} features while, at the same time, not significantly affecting the magnetic field topology around the affected regions. 
   }


   \maketitle
%
\nolinenumbers

\section{Introduction}
\label{sec:introduction}
With our growing reliance on technology sensitive to space weather effects, the need for accurate space weather modelling has also increased. Traditional frameworks used for such modelling and forecasting generally work with a version of the Wang-Sheeley-Arge model \citep{Arge2003} to derive the conditions in the solar corona, as for instance, EUHFORIA (EUropean Heliosphere FORecasting Information Asset) \citep{Pomoell18}. Such models are, however, only semi-empirical and often cannot capture the full complexity of the corona, and thus more elaborate physics-based models might be required (see, e.g. \cite{Samara2021}). The need to improve the accuracy of the corona boundary prescription and to gain insights into the more fundamental coronal physics has led to the development of more advanced full 3D magnetohydrodynamic (MHD) global coronal models (GCM). Examples of such models include the Wind-Predict code \citep{Reville_2015, Parenti2022}, the MAS (Magnetohydrodynamics Around a Sphere) model \citep{Mikic1996, Mikic1999, Linker1999}, and the AWSoM model (Alfvén Wave Solar Model) \citep{vanderHolst2014, Gombosi2018,Shi2022}. This paper will focus on the rapidly converging COolfluid COrona uNstrUcTured (COCONUT) global corona model \citep{PerriLeitner2022}. 

However, the results of GCMs are only as good as their prescription. Arguably, the most important input into these models is the photospheric magnetogram, or the magnetic map, which defines the electromagnetic features that will form in the solution domain. To consult how the different types of available magnetogram products perform in COCONUT global corona simulations and how their post-processing influences the resolved features, we refer the reader to \citet{Perri2023}, \citet{Kuzma2023} and \citet{Brchnelova2023Beta}. 

These models, however, also require the prescription of the thermodynamic values that are representative of the lower coronal conditions at the inner boundary. This task is challenging since this kind of real-time global information is generally unavailable from observations. For this reason, general and often homogeneous global profiles are usually prescribed for the temperature, density and velocity \citep{PerriLeitner2022}. However, assuming the same (or at least similar) thermodynamic conditions for quiet sun (QS) regions as in active regions or coronal holes is physically inappropriate{, as can be seen, for instance, in Figures 5 and 6 in the work of \citet{Bourdin2020}.}

As was demonstrated by \citet{Kuzma2023} and further discussed by \citet{Brchnelova2023Beta}, in COCONUT, assuming nonphysical thermodynamic boundary conditions, especially in the cases of solar maxima, may lead to a formation of {unexpected} streams in the domain. This is understandable as, in active regions, one could expect much higher densities and pressures than in quiet regions and coronal holes, as can be deduced, among others, from the work of \citet{Doschek_1998}. Prescribing a thermal pressure value that is too low with respect to the background magnetic pressure (i.e. a too-low plasma $\beta$) then leads to disproportionally large electromagnetic contributions to the momentum and energy equations.

For example, if we prescribe a $|\mathbf{B}|_\text{max}$ of $50$ to $100\;$G in the corona above active regions (the magnitude of which corresponds well to the magnetic field strengths in active regions as predicted by \citet{AlissandrakisGary2021}) with our default boundary conditions for pressure ($0.00416\;$Pa), we obtain plasma $\beta$'s of $10^{-4}$ to $4 \cdot 10^{-4}$. This is one to two orders of magnitude below what would be expected at the altitude of our inner boundary ($\sim 10\;$Mm) according to \cite{Gary2001}. Defining a locally higher thermal pressure would thus be one of the solutions to improve the prescription. 
 
Some of the available models solve this challenge by initiating the global simulations in the lower atmospheric layers of the Sun, such as the AWSoM model that is initiated in the upper chromosphere with a temperature of $50000\;$K and ion number density of $2 \cdot 10^{16}\;$m$^{-3}$ (readers can refer to \citet{vanderHolst2014}). The plasma has space throughout the transition region to adjust its thermodynamics better to the prescribed electromagnetic conditions before expanding into the rarefied and strongly magnetised coronal domain. However, adding a sufficient radial resolution to capture the transition region significantly increases the computational time. It may thus be undesirable for models intended for operational use, such as COCONUT. 

In this paper, we attempt to remove the {inexplicable} features from the domain without manipulating the grid resolution. By analysis of the available observational data and numerical simulation results from models such as Bifrost \citep{Gudiksen2011}, we recognise that a potentially more physically suitable way of modifying the inner boundary may be through constraining the prescribed plasma $\beta$ and Alfvén speed that act as proxies to constraining the plasma pressure and density. {The reason that plasma pressure and density have been chosen specifically is the fact that in the COCONUT solver, both of these are primitive variables, and are thus straightforward to prescribe directly in the formulation of the boundary condition.}


Section~\ref{sec:methodology} presents the code set-up, the rationale behind constraining the boundary plasma $\beta$, and the Alfvén speed and the method in which this constraining is implemented. Section~\ref{sec:results} shows the results of this technique when applied to the case of the high-activity 2016 solar eclipse. The paper is concluded in Section~\ref{sec:conclusion}.

   \begin{figure*}
   \centering
   \includegraphics[width=0.9\textwidth]{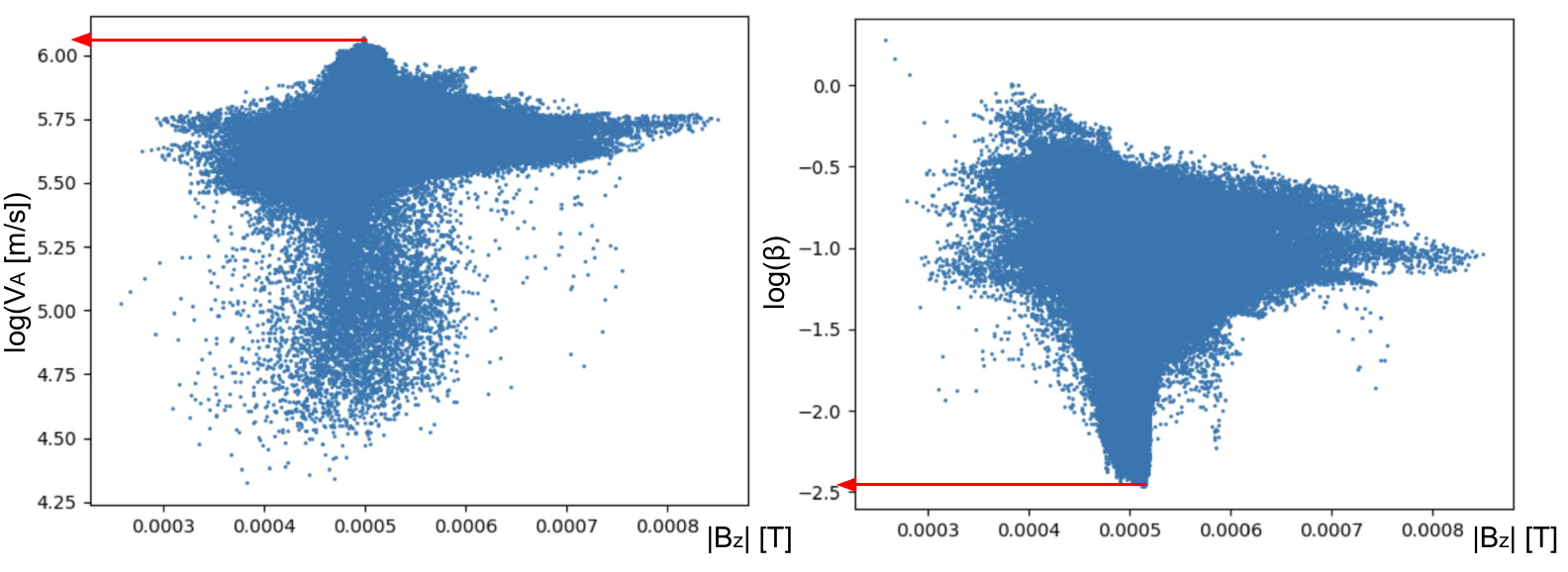}
   \caption{Shown are the logarithm of the Alfvén speed (left) and the logarithm of the plasma $\beta$ (right) as functions of the vertical magnetic field, as resolved in the Bifrost simulation ch024031\_by200bz005 \citep{Finley2022} at the height of the COCONUT GCM inner boundary, {at $\sim 10\;$Mm. The red arrows point at the maximum computed Alfvén speed (left) and the minimum resolved plasma $\beta$ (right) in this simulation.}}
        \label{fig:BifrostResults}%
    \end{figure*}
    
\section{Methodology}
\label{sec:methodology}

\subsection{The COCONUT global corona solver and setup}

In this work, we used the COCONUT solver \citep{PerriLeitner2022}, with the 6th level subdivided icosahedron-based grid as defined in \cite{Brchnelova2022a}. The default boundary conditions of COCONUT follow \cite{PerriLeitner2022} and assume a uniform inner boundary density $\rho_0$ and pressure $p_0$ of $1.67 \cdot 10^{-13}\;$kg/m$^3$ and $0.00416\;$Pa, respectively. A user-set velocity outflow is also prescribed aligned with the magnetic field  \citep{Brchnelova2022b}. 

The setup in use here considers the {compressible MHD} steady-state system (readers can refer to \citet{Baratashvili2024AAP}), which consists of the ideal-MHD equations with additional terms representing gravity, heat conduction (a collisionless approximation), optically thin radiative losses and a coronal heating function. {Heat conduction is defined according to \citet{Mikic1999} following \citet{Hollweg1978}, consisting of separate term for collisional (Spitzer-Härm, for $<10\;$R$_s$) and collisionless (Hollweg, for $> 10\;$R$_s$) conduction. The radiative loss is defined in the optically thin limit following \citet{Rosner1978}.} 

The form of the coronal heating approximation used here follows Equation~8 from \citet{Baratashvili2024AAP}, that is, coronal heating proportional to the magnetic field strength and following an envelope function, { $Q_{H} = H_0 \cdot |\mathbf{B}| \cdot e^{-\frac{r-R_s}{\lambda}}$, in which $H_0 = 4 \cdot 10^{-5}$ erg cm$^{-3}$ s$^{-1}$ G$^{-1}$ and $\lambda=0.7\;$R$_s$.}

We specifically focus on a case of the March 9, 2016, solar eclipse, corresponding to the Carrington rotation 2174. This case represents a solar maximum and was selected as it contains regions of strong magnetic fields of more than $50\;$G in the prescribed magnetic map when this map is obtained via standard processing. Herein the corresponding photospheric HMI magnetogram\footnote{\url{http://jsoc.stanford.edu/HMI/Magnetograms.html}} {\citep{Scherrer2012}} is post-processed using the spherical harmonics projection technique with $l_\text{max} = 25$; for the discussion of the suitable magnetogram products and the method of post-processing, we refer the reader to \citet{Perri2023}, \citet{Kuzma2023} and \citet{Brchnelova2023Beta}. The exact orientation was set such that it corresponds to the observation of the solar eclipse as seen from the Earth.

\subsection{Plasma $\beta$ and Alfvén speed constraints}
\label{subsec:contraints}

As mentioned in Section~\ref{sec:introduction}, the default homogeneous boundary conditions for density and pressure might be inaccurate, especially in regions outside QS conditions. The question thus turns to what the realistic ranges for these variables would be. {While we could focus on plasma density and pressure directly, we are interested in constraining those regions that have a stronger magnetic field. We would thus have to create expressions for these variables with the magnetic field dependence. The more straightforward alternative to this is, instead, to consider plasma $\beta$ and the Alfvén speed as proxies for the pressure and density as these already contain the magnetic field information in them.}

At the height of our inner boundary, which is $\sim 10\;$Mm, \citet{Gary2001} expects plasma $\beta$'s in the range between $10^{-3}$ to $2 \cdot 10^{-2}$. \citet{Iwai2014} estimated a plasma $\beta$ of $5.7 \cdot 10^{-4}$ to $7.6 \cdot 10^{-4}$ at the top of a post-flare loop at $4\;$Mm (which is slightly lower than what we assume) from satellite and radio observations. This observed range was also supported by \citet{Bourdin2013} and \citet{Bourdin2017} through their MHD model, giving ranges between $10^{-4}$ to $10^{-1}$ for active regions and $10^{-4}$ to $10^2$ for QS regions. The Alfvén speed in eight example coronal loops has been estimated via MHD seismology by \citet{Anfinogentov2019} to lie in the range between $8.66 \cdot 10^{5}\;$m/s to $1.17 \cdot 10^{6}\;$m/s. 

However, the amount of observational data is fairly limited, especially for the regions of coronal holes. For that reason, additionally, we used results of the  Bifrost code \citep{Gudiksen2011}, a state-of-the-art solver of the solar atmosphere that can resolve non-LTE radiative transfer. Specifically, the simulation data ch024031\_by200bz005, which were already previously analysed by \citet{Finley2022}, were probed to determine the existing $\beta$ and $V_A$ ranges resolved at the height that corresponds to the inner boundary of our GCM. The maximum values of $V_A$ were found to be $\sim 2\;$Mm/s and the minimum values of plasma $\beta$ of $\sim 0.003$, as can be seen in Figure~\ref{fig:BifrostResults}.

The values above are indirectly derived from observations, theory, or more detailed MHD simulations, so large uncertainties may exist in them. However, they do show that there is a rough consensus regarding the order of magnitude that would be expected at our inner boundary, which can be used to constrain our setup.

   \begin{figure*}
   \centering
   \includegraphics[width=0.9\textwidth]{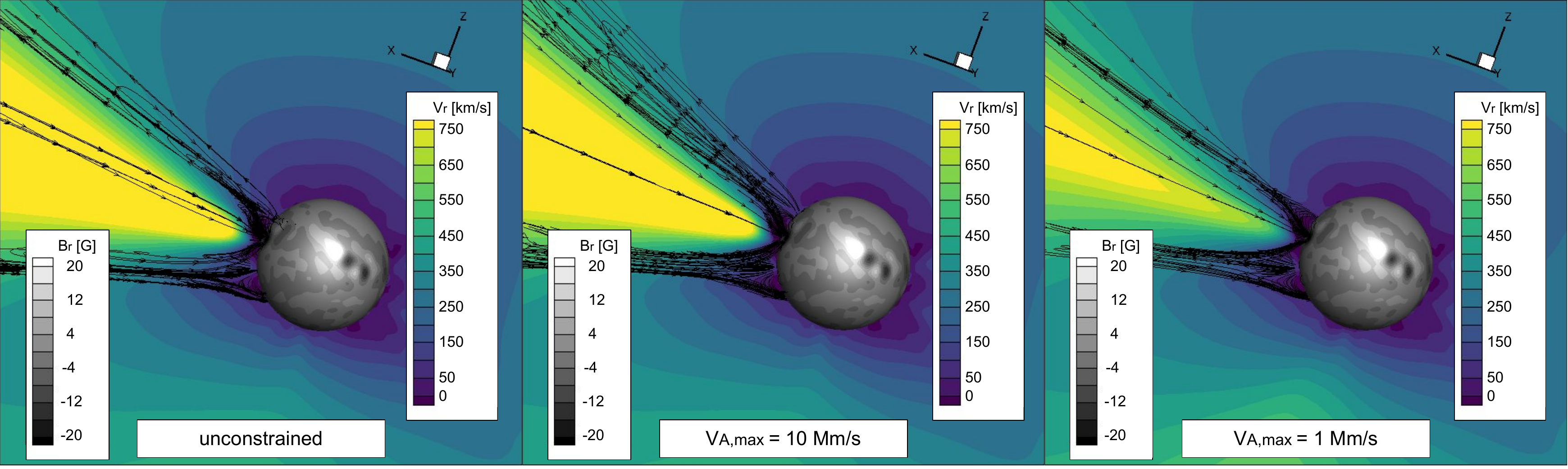}
   \caption{Shown are COCONUT results for the March 9, 2016, solar eclipse (CR2174), with maximum $V_A$ constraining. The inner surface shows the prescribed magnetic field, the contour plot of the radial speed (demonstrating the existence of {inexplicable} high-speed streams and their reduction thanks to the constraining), and the magnetic field lines over and around the active region, causing the high-speed streams plotted.}
        \label{fig:Bifrostconstraining_baseline_results_VA}%
    \end{figure*}

   \begin{figure*}
   \centering
   \includegraphics[width=0.9\textwidth]{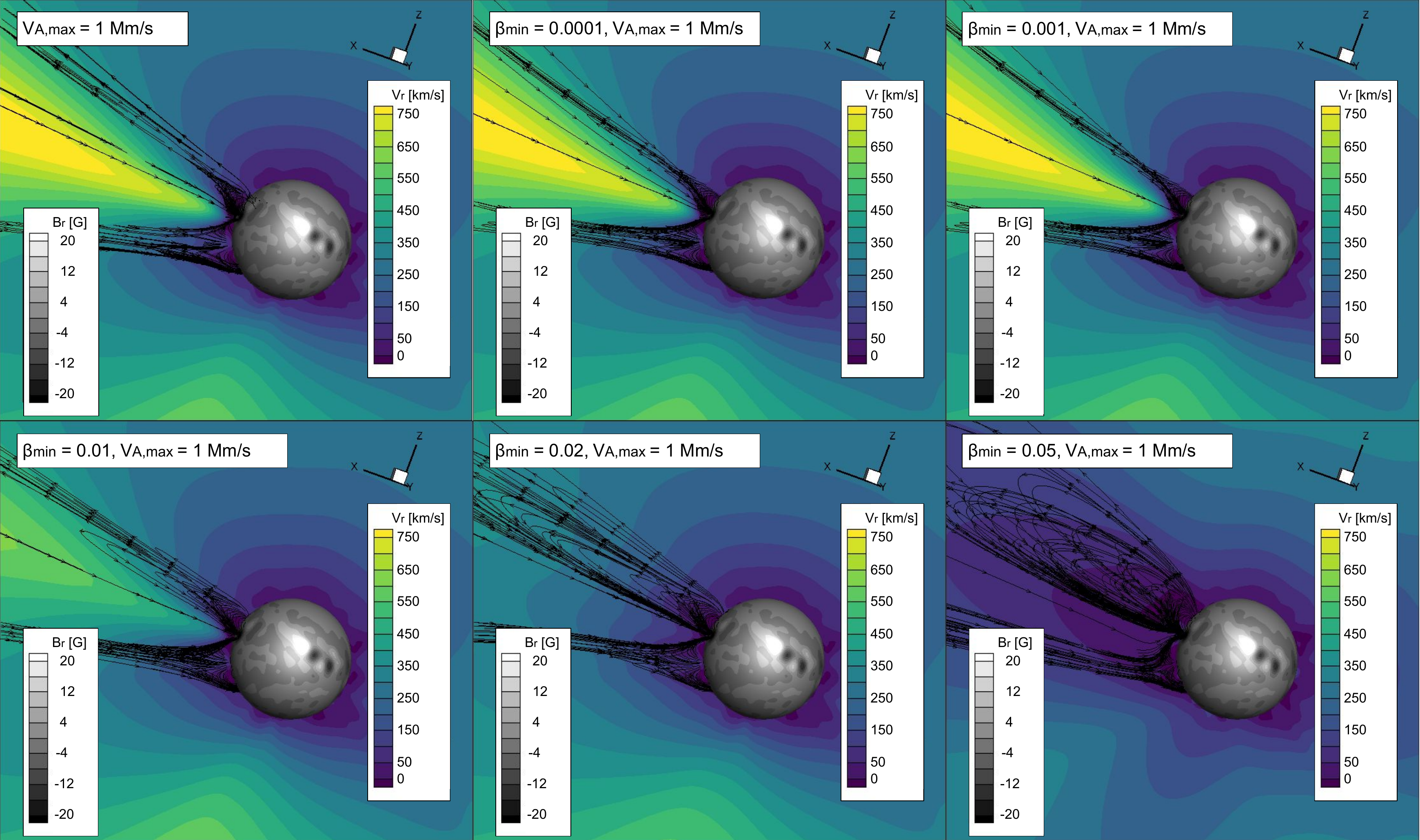}
   \caption{Shown are COCONUT results for the March 9, 2016, solar eclipse (CR2174), with minimum-plasma-$\beta$ constraining. The inner surface shows the prescribed magnetic field, the contour plot of the radial speed (demonstrating the existence of {inexplicable} high-speed streams and their removal thanks to the constraining), and the magnetic field lines over and around the active region, causing the high-speed streams plotted.}
        \label{fig:Bifrostconstraining_baseline_results_beta}%
    \end{figure*}

   \begin{figure*}
   \centering
   \includegraphics[width=0.9\textwidth]{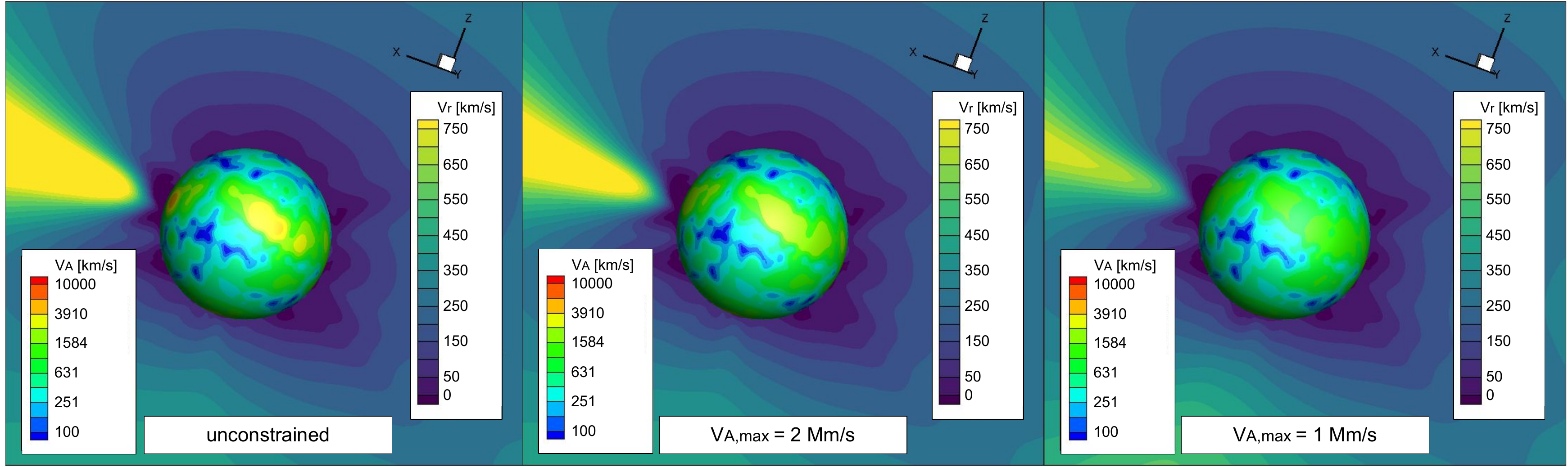}
   \caption{For the March 9, 2016 solar eclipse (CR2174), the changes to the prescribed Alfvén speed $V_A$ on the inner boundary are shown for three cases as a consequence of $V_{A,\text{max}}$ constraining. The highest $V_{A,\text{max}}$ solution on the left looked the same as the unconstrained one.}
        \label{fig:Bifrostconstraining_vA}%
    \end{figure*}

   \begin{figure*}
   \centering
   \includegraphics[width=0.9\textwidth]{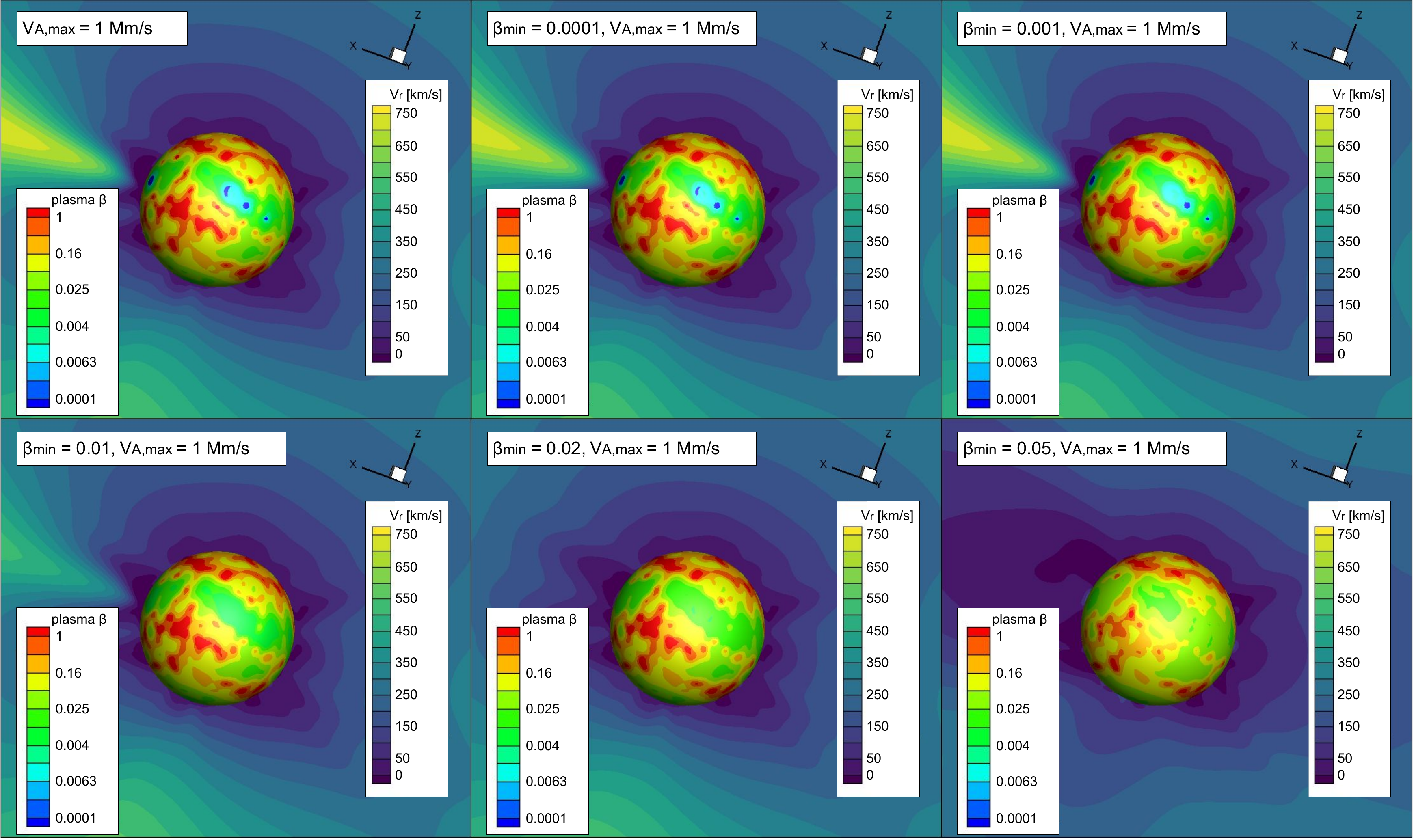}
   \caption{For the March 9, 2016, solar eclipse (CR2174), the changes to the prescribed plasma $\beta$ on the inner boundary are shown for six cases as a consequence of $\beta_\text{min}$ constraining. The lowest $\beta_\text{min}$ solution on the left looked the same as the one unconstrained with $\beta_\text{min}$, only with $V_{A,\text{max}}$.}
        \label{fig:Bifrostconstraining_beta}%
    \end{figure*}

   \begin{figure*}
   \centering
   \includegraphics[width=0.8\textwidth]{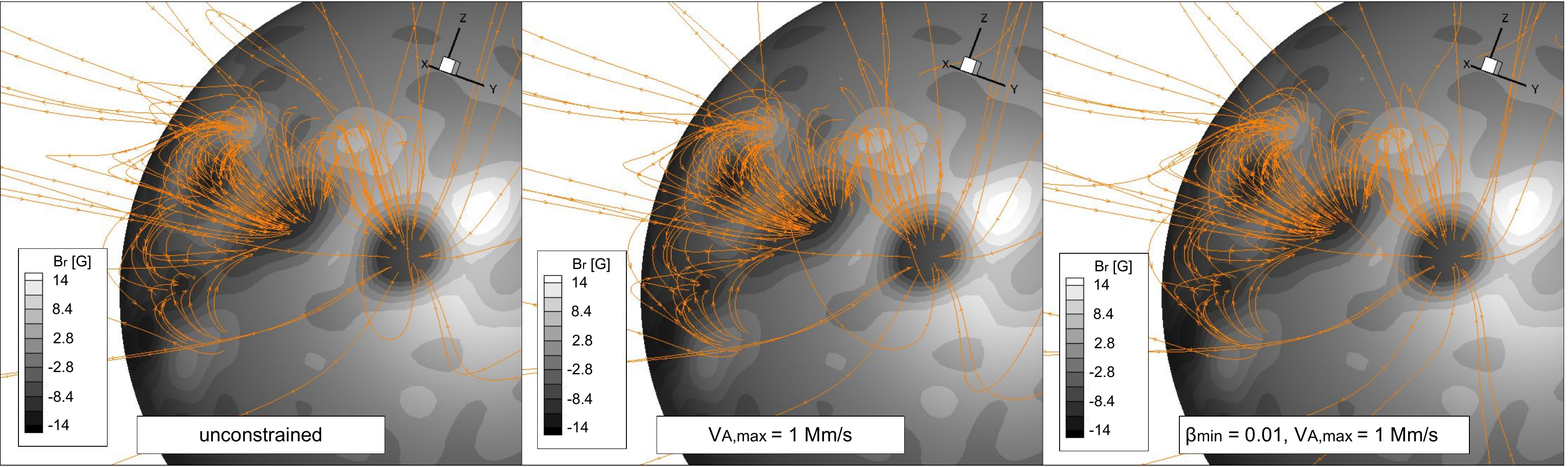}
   \caption{For the March 9, 2016 solar eclipse (CR2174), the detailed magnetic field lines are plotted over and around the strong active region causing the {inexplicable} high-speed streams for three levels of constraining (effectively no constraining on the left, $V_{A,\text{max}}=10^{6}\;$m/s in the middle and $V_{A,\text{max}}=10^{6}\;$m/s with  $\beta_\text{min}=10^{-2}$ constraining on the right).}
        \label{fig:Bifrostconstraining_lines}%
    \end{figure*}
    
\subsection{Numerical boundary condition constraining}

From the numerical standpoint, to constrain the plasma $\beta_\text{min}$ and $V_{A, \text{max}}$ to the values presented above, it is inappropriate to use {a non-steady step function} since such a function has a discontinuous first derivative {which triggered numerical convergence issues in our tests}. A smooth transition function should thus be designed between the homogeneous $p_0$ and $\rho_0$ values and the values computed from the constraints $p^\prime$ and $\rho^\prime$. This was achieved by implementing a double-sided hyperbolic tangent profile. For pressure, determined from $\beta_\text{min}$, we prescribed

\begin{equation}
	p^\prime = \zeta_\text{tan} p_\text{mag} \beta_\text{min} + p_0 (1 - \zeta_\text{tan}),
\end{equation}

in which $p_\text{mag}$ is the magnetic pressure computed from the prescribed magnetic field. The double-sided hyperbolic tangent transition factor $\zeta_\text{tan}$ is given by

\begin{equation}
	\zeta_\text{tan} = \frac{1}{2} + \frac{1}{2} \tanh{(\pi \Delta_\text{tan})} \quad \text{with} \quad \Delta_\text{tan} = \frac{\beta_\text{min} - \beta}{d_\text{tan}}.
\end{equation}

The term $\Delta$ is positive if $\beta < \beta_\text{min}$ (such that $\zeta_\text{tan} \to 1$) and negative if $\beta > \beta_\text{min}$ ($\zeta_\text{tan} \to 0$). The constant $d_\text{tan}$ defines the range of $\beta$ over which the transition occurs. If $d_\text{tan}$ is too small, convergence issues might occur (just like with a step function), whereas if $d_\text{tan}$ is too large, the constraining is no longer sharp and accurate. Based on the results of our numerical experiments, $d_\text{tan}$ was set to 10\% of $\beta_\text{min}$. For density, constrained by the maximum Alfvén speed, we then have

\begin{equation}
	\rho^\prime  = \zeta_\text{tan} \frac{|\textbf{B}|^2}{V_{A,\text{max}}^2 \mu_0}  + \rho_0 (1 - \zeta_\text{tan}),
\end{equation}

with

\begin{equation}
	\zeta_\text{tan} = \frac{1}{2} + \frac{1}{2} \tanh{(\pi \Delta_\text{tan})}, \quad \text{and} \quad \Delta_\text{tan} = \frac{V_A - V_{A, \text{max}}}{d_\text{tan}}.
\end{equation}

Based on the results of numerical experiments, we again set $d_\text{tan} \sim$ 10\% of $V_{A, \text{max}}$. Here, $p^\prime$ and $\rho^\prime$ are prescribed directly at the boundary. {In COCONUT, the size of the ghost cells, the state (i.e. the vector of primitive variables) of which is denoted by $g$, and the size of the innermost domain cells, denoted by $i$, is the same, and thus so is the distance from their centroids to the boundary. Thus, the ghost cell state values become defined by $p_\text{g} = 2 p^\prime - p_\text{i}$ and $\rho_\text{g} = 2 \rho^\prime  - \rho_\text{i}$ such that we achieve the values $p^\prime$ and $\rho^\prime$ exactly on the boundary.}

{Of course, the hyperbolic tangent profile only tends to the final values and never actually reaches them. However, thanks to the selection of such a small distance parameter, $d_\text{tan} \sim 10\%$ of $V_{A, \text{max}}$ and $\beta_\text{min}$, already, for instance, $10\%$ below and above the limit values of $\beta$ and $V_{A}$, the prescribed boundary pressure and density are within just 0.2\% percent of their target states. Given the other inherent uncertainties and inaccuracies in our boundary formulation (for instance, the magnetic field and the density field, as has been described in \citet{Brchnelova2023Beta} and \citet{Perri2023}, and the only first-order accurate interpolation used to prescribe the boundary values), such offsets are negligible. If needed, however, the value of $d_\text{tan}$ can be reduced further, though this could come at a possibly increased computational cost.}

\section{Results}
\label{sec:results}

The results were first obtained for an unconstrained simulation and then by constraining for several values of $V_{A,\text{max}}$ and $\beta_\text{min}$. The unconstrained simulation, as expected, produced a {high-speed} stream of speeds higher than $1\;$Mm/s; as can be seen in Figure~\ref{fig:Bifrostconstraining_baseline_results_VA} on the left. {This stream is unexpected given the underlying magnetic structure, and resolved to be much hotter than what would be realistic for any such kind of structure in the solar corona (here, with well over $30\;$MK)}. The width of this stream was progressively reduced via decreasing $V_{A,\text{max}}$, first to $V_{A,\text{max}}=10^{7}\;$m/s in the middle and then down to $V_{A,\text{max}}=10^{6}\;$m/s on the right, with the latter value roughly corresponding to what would be expected from the discussion in Subsection~\ref{subsec:contraints}.

From then on, to further reduce the strength of the stream, the $\beta$ constraining was included with the results shown in Figure~\ref{fig:Bifrostconstraining_baseline_results_beta} for the values of $\beta_\text{min}=10^{-4}$, $10^{-3}$, $10^{-2}$, $2 \cdot 10^{-2}$, and $5 \cdot 10^{-2}$. For the more realistic $\beta_\text{min}$ values ($10^{-3}$ and $10^{-2}$ according to \citet{Gary2001}), this stream first reaches lower speeds until it is mainly removed {in the range between $10^{-2} \leq \beta_\text{min} \leq 2 \cdot 10^{-2}$}, without a significant alteration to the surrounding electromagnetic structures. For $\beta_\text{min}$ values that are too high ($2 \cdot 10^{-2}$, $5 \cdot 10^{-2}$), the flow field and the magnetic field lines become highly deformed.

Figure~\ref{fig:Bifrostconstraining_vA} depicts, for three of the selected $V_{A,\text{max}}$ cases, how the prescribed $V_A$ on the surface of the inner boundary changes as a result of the numerical constraining (the surface $V_{A}$ looking almost identical for $V_{A,\text{max}}=10^{7}\;$m/s and the unconstrained case). {By comparing the inner boundary surfaces visible here with those in Figure~\ref{fig:Bifrostconstraining_baseline_results_VA}, it is clear that the locations of the highest $V_A$ correspond to the regions of the strongest magnetic field, as would be expected. It is also these regions that are affected by limiting $V_{A,\text{max}}$, with these high values being progressively removed in Figure~\ref{fig:Bifrostconstraining_vA} from left to the right, and with the rest of the inner boundary $V_A$ surface being unaffected.} The same is shown in Figure~\ref{fig:Bifrostconstraining_beta} for plasma $\beta$, with the $\beta$-unconstrained and $\beta_\text{min}=10^{-4}$ again looking very similar. {As was the case of $V_A$, the regions of the smallest plasma $\beta$ correspond to the those where the magnetic field is the strongest, and it is only these regions where the $\beta$ values are affected as we increase $\beta_\text{min}$ from $10^{-4}$ at the top, left to $\beta_\text{min} \geq 10^{-2}$ in the bottom row.}

{With the selection on the limiting $\beta$ and $V_A$, one can also analyse the effects that this technique has on the assumed temperature (which is not a primitive variable in the solver, and so it must be derived from the pressure and density, with $T \propto p / \rho$). Reducing $V_{A,\text{max}}$ means increasing the surface density, lowering the corresponding temperature. In contrast, increasing $\beta_\text{min}$ means increasing the surface pressure and thus the temperature. As a result, controlling these two parameters together may be used to design the desired target temperature in the constrained regions. An example of such a calculation for a $5\;$G region (that corresponds to a boundary magnetic field that is very typical in our maxima simulations) with a variety of $V_{A,\text{max}}$ and $\beta_\text{min}$ values is shown in Table \ref{tab:temperature}.}

\begin{table}
\caption{Temperatures for a variety of $V_{A,\text{max}}$ and $\beta_\text{min}$.}
\begin{tabular}{l|l|l}
\hline \hline
plasma $\beta_\text{min}$ & $V_{A,\text{max}}$ [m/s]    & $T$ [K]     \\ \hline
$10^{-4}$ & $10^7$  & $3.05 \cdot 10^7$ \\ 
$10^{-3}$ & $10^7$ &$3.05 \cdot 10^8$  \\
$10^{-3}$ & $2 \cdot 10^6$  & $1.22 \cdot 10^7$ \\
$10^{-3}$ & $10^6$  & $3.05 \cdot 10^6$  \\
$10^{-2}$ & $10^6$ & $3.05 \cdot 10^7$ \\ \hline \hline
\end{tabular}
\tablefoot{For a variety of values of $V_{A,\text{max}}$ and $\beta_\text{min}$, it is shown how these can be used in combination to reach a desired temperature in the constrained region (here for an example region with an assumed magnetic field strength of $5\;$G).}
\label{tab:temperature}
\end{table}


Notably, with this approach, the magnetic field lines resolved over these affected regions remain mostly unaltered for the physically justifiable constraining levels. In Figure~\ref{fig:Bifrostconstraining_lines}, the magnetic field lines over and around the region causing the {inexplicable} stream are shown in orange. The left-most plot corresponds to an effectively unconstrained simulation, and the other plots then show the field lines with $V_A$ constraining (middle) and $V_A$ plus $\beta$ constraining (right). While the magnetic field lines over the active region are somewhat sharper in the left-most case, the magnetic connectivity and general structure of the field lines remain very similar.

\section{Discussion and conclusion}
\label{sec:conclusion}

In this paper, we have developed and demonstrated a method to alter the pressure and density boundary condition formulations of the global coronal model COCONUT \citep{PerriLeitner2022} to remove {inexplicable} features in the domain. These features stem from the fact that in the default model setup, the density and pressure are assumed to be homogeneous all across the coronal boundary surface, which is especially inaccurate in regions with stronger magnetic fields. Inappropriate density and pressure prescriptions then lead to nonphysically large Alfvén speeds and low plasma $\beta$ values, affecting the plasma dynamics. 

By analysing literature findings and the data from the solar atmospheric code Bifrost \citep{Gudiksen2011}, realistic orders of magnitude of the minimum and maximum constraints on the boundary $\beta$ and Alfvén speed were identified, with these parameters acting as proxies to the boundary pressure (from plasma $\beta$) and density (from Alfvén speed). To ensure a smooth transition and avoid convergence issues, a double-sided hyperbolic transition profile was employed for this constraining. 

Tests conducted on the 2016 solar eclipse case (March 9, CR 2174) demonstrated that $V_{A,\text{max}}$ constraining with the values that would be expected in the lower corona ($\sim 10^6$) contributed to the reduction of the radial width of the {inexplicable} features. Further reduction was achieved via $\beta_\text{min}$-constraining, with the most effective values being $10^{-3} < \beta_\text{min} \leq 1 \cdot 10^{-2}$, above which the $\beta$ became nonphysically high and led to a distortion of the flow and magnetic fields. {The specific values of $V_{A,\text{max}}$ and $\beta_\text{min}$ can be chosen to achieve a specific target temperature in the regions undergoing limiting, depending on the strength of the magnetic field.} It was also confirmed that for a reasonable selection of $V_{A,\text{max}}$ and $\beta_\text{min}$, beyond the removal of the {inexplicable} feature, the shape and the configuration of the magnetic field lines above the constrained regions were not significantly affected by the adopted method.

Thanks to missions such as the Solar Orbiter and the Parker Solar Probe, we can hopefully soon acquire new high-resolution observations allowing us to prescribe solar coronal conditions more accurately. Until that is the case, however, a technique such as the one presented in this paper might serve as a partial remedy to constrain global coronal models.

\begin{acknowledgements}
      This research was supported by the Research Council of Norway through its Centres of Excellence scheme, project number 262622, and through grants of computing time from the Programme for Supercomputing. This work has been further granted by the AFOSR basic research initiative project FA9550-18-1-0093. The project has also received funding from the European Union’s Horizon 2020 research and innovation programme under grant agreement No 870405 (EUHFORIA 2.0). These results were also obtained in the framework of the projects C16/24/010  (C1 project Internal Funds KU Leuven), G0B5823N and G002523N (WEAVE)  (FWO-Vlaanderen), V461823N (FWO-Vlaanderen), 4000134474 (SIDC Data Exploitation, ESA Prodex), and Belspo project B2/191/P1/SWiM. The resources and services used in this work were provided by the VSC (Flemish Supercomputer Centre), funded by the Research Foundation - Flanders (FWO) and the Flemish Government.
\end{acknowledgements}

\bibliographystyle{aa} 
\bibliography{biblio.bib}

\end{document}